\documentclass[numreferences]{kluwer}    

\usepackage{amssymb}
\usepackage{euscript}
\usepackage{graphicx}

\newcommand{\ac}[1]{c:\Sigma\times\Sigma^{\leq{#1}}\rightarrow\Delta^{+}}
\newcommand{\eac}{\overline{c}:\Sigma^{*}\rightarrow\Delta^{*}}
\newcommand{\sstring}[1]{\sigma_{1}\sigma_{2}\ldots\sigma_{#1}}

\newtheorem{definition}{Definition}
\newtheorem{example}{Example}
\newtheorem{theorem}{Theorem}

\begin{document}                                                                                   
\begin{article}
\begin{opening}         
\title{Modelling the EAH Data Compression Algorithm using Graph Theory
	\thanks{The research reported in this paper was partially supported by CNCSIS grant 632/2004.}} 
\author{Drago\c s N. \surname{Trinc\u a}\email{dragost@infoiasi.ro}}  
\runningauthor{Drago\c s N. Trinc\u a}
\runningtitle{Modelling the EAH Data Compression Algorithm using the Graph Theory}
\institute{Faculty of Computer Science, ``Al.I.Cuza'' University, 700483 Iasi, Romania}

\begin{abstract}
	Adaptive codes associate variable-length codewords
	to symbols being encoded depending on the previous symbols in the input data string.
	This class of codes has been introduced in \cite{t:1} as a new class of non-standard variable-length codes.
	New algorithms for data compression,
	based on adaptive codes of order one, have been presented in \cite{t:2},
	where we have behaviorally shown that for a large class of input data strings, these algorithms
	substantially outperform the Lempel-Ziv universal data compression algorithm \cite{zl:1}.
	\textsf{EAH} has been introduced in \cite{t:3}, as an improved generalization of these algorithms.
	In this paper, we present a translation of the \textsf{EAH} algorithm into the graph theory.
\end{abstract}

\end{opening}           

\section{Introduction}
	New algorithms for data compression, based on adaptive codes of order one \cite{t:1},
	have been presented
	in \cite{t:2}, where we have behaviorally shown that for a large class of input data strings, these
	algorithms substantially outperform the well-known Lempel-Ziv universal data compression algorithm
	\cite{zl:1}.
\newline\indent
	\textsf{EAH} (Encoder based both on Adaptive codes and the Huf\mbox{}fman algorithm)
	has been proposed in \cite{t:3}, as an improved and generalized version
	of the algorithms presented in \cite{t:2}. The work carried out so far \cite{t:2,t:3}
	has behaviorally proved that \textsf{EAH} is a very promising data compression algorithm,
	as one can remark again in the examples presented in the following sections. In this paper, we
	translate the \textsf{EAH} encoder, as well as some of the results obtained so far, into the graph theory.
\newline\indent
	In the remainder of this section, we recall the basic notions and notations used
	throughout the paper.
	We denote by $|S|$ the \textit{cardinality} of the set $S$; if $x$ is a string of
	f\mbox{}inite length, then $|x|$ denotes the length of $x$.
	The \textit{empty word} is denoted by $\lambda$.
	Let us
	denote by $\Delta^{*}$ the set
	$\bigcup_{n=0}^{\infty}\Delta^{n}$ and
	by $\Delta^{+}$ the set
	$\bigcup_{n=1}^{\infty}\Delta^{n}$.
	Also, denote by $\Delta^{\leq n}$ the set
	$\bigcup_{i=0}^{n}\Delta^{i}$
	and by $\Delta^{\geq n}$ the set
	$\bigcup_{i=n}^{\infty}\Delta^{i}$, where $\Delta^{0}$ denotes the set $\{\lambda\}$.
	$\mathbb{N}$ denotes the set of natural numbers, and $\mathbb{N}^{*}=\mathbb{N}-\{0\}$.
\newline\indent
	Let $X$ be a f\mbox{}inite and nonempty subset of $\Delta^{+}$, and $w\in\Delta^{+}$.
	A \textit{decomposition of w} over $X$ is any sequence of words
	$u_{1}, u_{2}, \ldots, u_{h}$ with $u_{i}\in X$, $1\leq i\leq h$, 
	such that $w=u_{1}u_{2}\ldots u_{h}$.
	A \textit{code} over $\Delta$ is any nonempty set $C\subseteq\Delta^{+}$ such
	that each word $w\in\Delta^{+}$ has at most one decomposition over $C$.
	A \textit{pref\mbox{}ix code} over $\Delta$ is any code $C$ over $\Delta$ such that
	no word in $C$ is proper pref\mbox{}ix of another word in $C$.

\section{Adaptive Codes: A Short Review}
	The aim of this section is to brief\mbox{}ly review some basic
	def\mbox{}initions, results, and notations directly related to adaptive codes.
\begin{definition}
	Let $\Sigma$, $\Delta$ be alphabets. A function
	$\ac{n}$, $n\geq{1}$, is called \textup{adaptive code of order $n$} if its unique
	homomorphic extension $\eac$, def\mbox{}ined by:
\begin{itemize}[$\bullet$]
\item $\overline{c}(\lambda)=\lambda$ 
\item $\overline{c}(\sstring{m})=$
	$c(\sigma_{1},\lambda)$
	$c(\sigma_{2},\sigma_{1})$
	$\ldots$
	$c(\sigma_{n-1},\sstring{n-2})$
	\newline\vspace{0pt}\hspace{4.2pt}
	$c(\sigma_{n},\sstring{n-1})$
	$c(\sigma_{n+1},\sstring{n})$
	$c(\sigma_{n+2},\sigma_{2}\sigma_{3}\ldots\sigma_{n+1})$
	\newline\vspace{0pt}\hspace{4.2pt}
	$c(\sigma_{n+3},\sigma_{3}\sigma_{4}\ldots\sigma_{n+2})\ldots$
	$c(\sigma_{m},\sigma_{m-n}\sigma_{m-n+1}\ldots\sigma_{m-1})$
\end{itemize}
	for all $\sstring{m}\in\Sigma^{+}$, is injective.
\end{definition}
\hspace{0pt}\indent
	As specif\mbox{}ied by the def\mbox{}inition above, an adaptive code of order $n$
	associates variable-length
	codewords to symbols being encoded depending on the previous $n$ symbols in the input data string. 
\begin{example}
	Let us consider $\Sigma=\{a,b,c\}$ and $\Delta=\{0,1\}$ two alphabets, and
	$\ac{2}$ a function constructed by the following table.
	One can easily verify that $\overline{c}$ is injective, and according to DEF. 1, $c$ is
	an adaptive code of order two. 
\begin{table}[h]
\caption{An adaptive code of order two.}
\begin{tabular}{llllllllllllll}
\hline
$\Sigma\backslash\Sigma^{\leq{2}}$	& a  & b	& c	& aa	& ab & ac & ba & bb & bc & ca & cb & cc & $\lambda$\\ \hline
a 					& 00 & 11 	& 10 	& 00 	& 11 & 10 & 01 & 10 & 11 & 11 & 11 & 00 & 00	   \\ 
b					& 10 & 00	& 11	& 11	& 01 & 00 & 00 & 11 & 01 & 10 & 00 & 10 & 11       \\ 
c					& 11 & 01	& 00	& 10	& 00 & 11 & 11 & 00 & 00 & 00 & 10 & 11 & 10	   \\ \hline
\end{tabular}	
\end{table}
\newline
	Let $x=abacca\in\Sigma^{+}$ be an input data string. 
	Using the def\mbox{}inition above, we encode $x$ by:
\begin{center}
	$\overline{c}(x)=c(a,\lambda)c(b,a)c(a,ab)c(c,ba)c(c,ac)c(a,cc)=001011111100$.
\end{center}
\end{example}
\hspace{0pt}\indent
	Let $\ac{n}$ be an adaptive code of order $n$, $n\geq{1}$. We denote by
	$\mathcal{C}_{c, \sigma_{1}\sigma_{2}\ldots\sigma_{h}}$ the set
	$\{c(\sigma,\sigma_{1}\sigma_{2}\ldots\sigma_{h}) \mid \sigma\in\Sigma\}$,
	for all $\sigma_{1}\sigma_{2}\ldots\sigma_{h}\in\Sigma^{\leq{n}}-\{\lambda\}$,
	and by $\mathcal{C}_{c, \lambda}$ the set $\{c(\sigma,\lambda) \mid \sigma\in\Sigma\}$.
	We write $\mathcal{C}_{\sigma_{1}\sigma_{2}\ldots\sigma_{h}}$ instead of
	$\mathcal{C}_{c, \sigma_{1}\sigma_{2}\ldots\sigma_{h}}$,
	and $\mathcal{C}_{\lambda}$ instead of $\mathcal{C}_{c, \lambda}$
	whenever there is no confusion.
\newline\indent
	Let us denote by $\mathcal{AC}(\Sigma,\Delta,n)$ the set
	$\{\ac{n} \mid c$ is an adaptive code of order $n\}$.
	The proof of the following theorem can be found in \cite{t:1}.
\begin{theorem}
	Let $\Sigma$ and $\Delta$ be two alphabets, and $\ac{n}$ a function.
	If $\mathcal{C}_{u}$ is pref\mbox{}ix code, for all $u\in\Sigma^{\leq{n}}$, then
	$c\in{\mathcal{AC}(\Sigma,\Delta,n)}$.
\end{theorem}

\section{Modelling the EAH Encoder using the Graph Theory}
	This section focuses on modelling some of the results obtained so far, as well as
	the \textsf{EAH} data compression algorithm \cite{t:3}, using the graph theory. 
\newline
\begin{definition}
	Let $\Sigma$ be an alphabet, and $w=u_{1}u_{2}\ldots{u_{h}}\in\Sigma^{+}$, where
	$u_{i}\in\Sigma$ for all $i$, $1\leq{i}\leq{h}$.
	The \textup{adaptive graph of order $n$} associated to $w$ is
	a 3-tuple $G_{n}(w)=(V,E,f)$, where $f:E\rightarrow{\mathbb{N}\times\{0,1\}^{*}}$ is a function,
	and $(V,E)$ is a directed graph def\mbox{}ined as below.
\begin{itemize}[$\bullet$]
\item $V=Vbase\cup{Vaux}$.
\item $E=Ebase\cup{Eaux}$.
\item $Vbase=
	\left\{
		\begin{array}{ll}
		\emptyset					& \textrm{if $h\leq{n}$.}	\\
		\{u_{j}u_{j+1}\ldots{u_{j+n-1}}\mid 1\leq{j}\leq{h-n}\} 
								&			 		\\
		\hspace{64pt}\cup\{u_{j}\mid n+1\leq{j}\leq{h}\}
								& \textrm{if $h>n$.}		\\
	\end{array}
	\right.$
\item $Vaux=
	\left\{
		\begin{array}{ll}
		\emptyset					& \textrm{if $n\geq{2}$.}	\\
		\{u_{j}\_aux\mid j\leq{h-1}$ and $u_{j}=u_{j+1}\} 
								& \textrm{if $n=1$.}		\\
	\end{array}
	\right.$
\item $Ebase=
	\left\{
		\begin{array}{ll}
		\emptyset					& \textrm{if $h\leq{n}$.}	\\
		\{(u_{j}\ldots{u_{j+n-1}},u_{j+n})\mid j\geq{1}\} 
								& \textrm{if $h>n$ and $n\geq{2}$.}		\\
		\{(u_{j},u_{j+1}) \mid u_{j}\neq{u_{j+1}} \}
								& \textrm{if $h>n$ and $n=1$.}		\\							
	\end{array}
	\right.$
\item $Eaux=
	\left\{
		\begin{array}{ll}
		\emptyset					& \textrm{if $h\leq{n}$.}	\\
		\{(u_{j},u_{j}\_aux)\mid u_{j}=u_{j+1}\}
								&					\\
		\cup\{(u_{j}\_aux,u_{j})\mid u_{j}=u_{j+1}\}
								& \textrm{if $h>n=1$.}	\\
		\{(u_{j},u_{j-n+1}\ldots{u_{j}})\mid n+1\leq{j}\leq{h-1}\}
								& \textrm{if $h>n\geq{2}$.}		\\
	\end{array}
	\right.$
\end{itemize}
	For all $e\in{E}$, if $f(e)=(x,y)$ then $x$ is def\mbox{}ined as follows.
\begin{displaymath}
	x=
	\left\{
	\begin{array}{ll}
		|\{i \mid u_{i}\ldots{u_{i+n}}=u_{j}\ldots{u_{j+n}}\}|
				& \textrm{if $e=(u_{j}\ldots{u_{j+n-1}},u_{j+n})$.}\\
		|\{i \mid u_{i}=u_{i+1}=u_{j}\}|
				& \textrm{if $e=(u_{j},u_{j}\_aux)$.}\\
		0
				& \textrm{if $e=(u_{j}\_aux,u_{j})$.}\\
		0
				& \textrm{if $e=(u_{j},u_{j-n+1}\ldots{u_{j}})$.}\\
	\end{array}
	\right.
\end{displaymath}
\end{definition}
\hspace{0pt}\newline\newline\indent
	Let us introduce some basic notations used throughout the section, and then
	give a few examples of adaptive graphs. Let $\Sigma$ be an alphabet, $w\in\Sigma^{+}$,
	$G_{n}(w)=(V,E,f)$, and $x\in{V=Vbase\cup{Vaux}}$.
\begin{itemize}[$\bullet$]
\item $input_{b}^{G_{n}(w)}(x)=\{(y,x)\mid y\in{Vbase}\}$.
\item $input_{a}^{G_{n}(w)}(x)=\{(y,x)\mid y\in{Vaux}\}$.
\item $output_{b}^{G_{n}(w)}(x)=\{(x,y)\mid y\in{Vbase}\}$.
\item $output_{a}^{G_{n}(w)}(x)=\{(x,y)\mid y\in{Vaux}\}$.
\item $i_{b}^{G_{n}(w)}(x)=|input_{b}^{G_{n}(w)}(x)|$.
\item	$o_{b}^{G_{n}(w)}(x)=|output_{b}^{G_{n}(w)}(x)|$.
\item	$i_{a}^{G_{n}(w)}(x)=|input_{a}^{G_{n}(w)}(x)|$.
\item $o_{a}^{G_{n}(w)}(x)=|output_{a}^{G_{n}(w)}(x)|$.
\end{itemize}
\begin{example}
	Let $\Sigma=\{a,b,c,d\}$ be an alphabet, $w=abccdbbab\in\Sigma^{+}$,
	and $G_{1}(w)=(V,E,f)$ the adaptive graph of order one associated to $w$.
	It is easy to verify that we get the following results.
\begin{itemize}[$\bullet$]
\item $V=\{a,b,c,d,c\_aux,b\_aux\}$.
\item $E=\{(a,b),(b,a),(b,b\_aux),(b\_aux,b),(b,c),(c,c\_aux),(c,d),\newline(c\_aux,c),(d,b)\}$.
\end{itemize}
\hspace{0pt}\indent
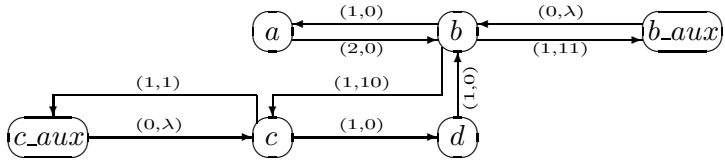
\begin{figure}[h]
\setlength{\unitlength}{1pt}
\begin{picture}(340,58)
	\put(132,10){\oval(15,15)}	
	\put(202,10){\oval(15,15)}
	\put(132,50){\oval(15,15)}
	\put(202,50){\oval(15,15)}
	\put(47,10){\oval(30,15)}
	\put(287,50){\oval(30,15)}

	\put(129,47){$a$}
	\put(199,47){$b$}
	\put(129,7){$c$}
	\put(199,7){$d$}
	\put(274,47){$b\_aux$}
	\put(34,7){$c\_aux$}

	\put(139.5,10){\vector(1,0){55}}
	\put(139.5,47){\vector(1,0){55}}
	\put(194.5,53){\vector(-1,0){55}}
	\put(209.5,47){\vector(1,0){62.3}}
	\put(271.8,53){\vector(-1,0){62.3}}
	\put(202,17.5){\vector(0,1){25}}
	\put(196,44){\line(0,-1){18}}
	\put(196,26){\line(-1,0){64}}
	\put(132,26){\vector(0,-1){8.2}}
	\put(62.5,10){\vector(1,0){62}}
	\put(126,15.3){\line(0,1){10.7}}
	\put(126,26){\line(-1,0){77}}
	\put(49,26){\vector(0,-1){8.2}}

	\put(157,56){$\scalebox{1}[0.7]{\scriptsize{(1,0)}}$}
	\put(157,42){$\scalebox{1}[0.7]{\scriptsize{(2,0)}}$}
	\put(155,29){$\scalebox{1}[0.7]{\scriptsize{(1,10)}}$}
	\put(157,13){$\scalebox{1}[0.7]{\scriptsize{(1,0)}}$}
	\put(80,29){$\scalebox{1}[0.7]{\scriptsize{(1,1)}}$}
	\put(80,13){$\scalebox{1}[0.7]{\scriptsize{(0,$\lambda$)}}$}
	\put(204,20){\rotatebox{90}{$\scalebox{1}[0.7]{\scriptsize{(1,0)}}$}}
	\put(232,56){$\scalebox{1}[0.7]{\scriptsize{(0,$\lambda$)}}$}
	\put(230,42){$\scalebox{1}[0.7]{\scriptsize{(1,11)}}$}
\end{picture}
\caption{$G_{1}(abccdbbab)$.}
\end{figure}
\end{example}
\hspace{0pt}\indent
	Let $U=(u_{1},u_{2},\ldots,u_{k})$ be a $k$-tuple.
	We denote by $U.i$ the $i$-th component of $U$,
	that is, $u_{i}=U.i$, for all $i$, $1\leq{i}\leq{k}$.
	The $0$-tuple is denoted by $()$. The length of a tuple $U$ is denoted by ${Len}(U)$.
	If $V=(v_{1},v_{2},\ldots,v_{b})$,
	$M=(m_{1},m_{2},\ldots,m_{r},U)$, $N=(n_{1},n_{2},\ldots,n_{s},V)$,
	$P=(p_{1},\ldots,p_{i-1},p_{i},p_{i+1},\ldots,p_{t})$ are tuples, and q is an element or a tuple,
	then we def\mbox{}ine
	$P\vartriangleleft{q}$, $P\vartriangleright{i}$, $U\vartriangle{V}$, and $M\lozenge{N}$ by:
\begin{itemize}[$\bullet$]
\item $P\vartriangleleft{q}=(p_{1},\ldots,p_{t},q)$
\item $P\vartriangleright{i}=(p_{1},\ldots,p_{i-1},p_{i+1},\ldots,p_{t})$
\item $U\vartriangle{V}=(u_{1},u_{2},\ldots,u_{k},v_{1},v_{2},\ldots,v_{b})$
\item $M\lozenge{N}=(m_{1}+n_{1},m_{2}+1,\ldots,m_{r}+1,n_{2}+1,\ldots,n_{s}+1,U\vartriangle{V})$
\end{itemize}
	where $m_{i}$, $n_{j}$ are integers, $1\leq{i}\leq{r}$ and $1\leq{j}\leq{s}$.
	Let us denote by $\textsf{Huffman}$
	the Huf\mbox{}fman algorithm, which gets as input a tuple $(f_{1},\ldots,f_{k})$ of integers, and
	returns a tuple $((c_{1},l_{1}),\ldots,(c_{k},l_{k}))$,
	where $c_{i}$ is the codeword associated to the symbol with the frequency $f_{i}$,
	and $l_{i}$ is the length of $c_{i}$ for all $i$, $1\leq{i}\leq{k}$.
\begin{figure}[h]
\setlength{\unitlength}{1pt}
\begin{picture}(335,150)
	\put(7,140){\footnotesize{$\mathsf{Huffman}$(tuple $\mathcal{F}=(f_{1},\ldots,f_{k})$, where $k\geq{1}$)}}
	\put(17,130){\footnotesize{$1.$ $\mathcal{L}:=((f_{1},0,(1)),\ldots,(f_{k},0,(k)));$
						Let $\mathcal{V}=(\lambda,\lambda,\ldots,\lambda)$ be a $k$-tuple;}}
	\put(17,120){\footnotesize{$2.$ if $k=1$ then $\mathcal{V}.1:=0;$}}
	\put(17,110){\footnotesize{$3.$ while $Len(\mathcal{L})>1$ do}}
	\put(27.6,100){\footnotesize{begin}}
	\put(37,90){\footnotesize{$3.1.$ Let $i<j$ be such that $1\leq{i,j}\leq{Len(\mathcal{L})}$
			and $\mathcal{L}.i.1$, $\mathcal{L}.j.1$ are}}
	\put(55,80){\footnotesize{the smallest elements
			of the set $\{\mathcal{L}.q.1 \mid 1\leq{q}\leq{Len(\mathcal{L})}\}$;}}
	\put(37,70){\footnotesize{$3.2.$
	$\mathit{First}:=\{\mathcal{L}.i.Len(\mathcal{L}.i).r\mid
			 1\leq{r}\leq{Len(\mathcal{L}.i.Len(\mathcal{L}.i))}\};$}}
	\put(37,60){\footnotesize{$3.3.$
$\mathit{Second}:=\{\mathcal{L}.j.Len(\mathcal{L}.j).r\mid 1\leq{r}\leq{Len(\mathcal{L}.j.Len(\mathcal{L}.j))}\};$}}
	\put(37,50){\footnotesize{$3.4.$ for each $x\in\mathit{First}$ do $\mathcal{V}.x:=0\cdot{\mathcal{V}.x};$}}
	\put(37,40){\footnotesize{$3.5.$ for each $x\in\mathit{Second}$ do $\mathcal{V}.x:=1\cdot{\mathcal{V}.x};$}}
	\put(37,30){\footnotesize{$3.6.$ $\mathcal{U}:=\mathcal{L}.i$ $\lozenge$ $\mathcal{L}.j;$
			$\mathcal{L}:=\mathcal{L}\vartriangleright{j};$
			$\mathcal{L}:=\mathcal{L}\vartriangleright{i};$
			$\mathcal{L}:=\mathcal{L}\vartriangleleft{\mathcal{U}};$}}
	\put(27.6,20){\footnotesize{end}}
	\put(17,10){\footnotesize{4.} return
			$((\mathcal{V}.1,|\mathcal{V}.1|),\ldots,(\mathcal{V}.k,|\mathcal{V}.k|));$}
	\put(2,150){\line(1,0){332}}
	\put(2,5){\line(1,0){332}}
	\put(2,150){\line(0,-1){145}}
	\put(334,150){\line(0,-1){145}}
\end{picture}
\caption{The Huf\mbox{}fman algorithm.}
\end{figure}
\hspace{0pt}\newline\indent
	Let $\Sigma=\{\sigma_{0},\sigma_{1},\ldots,\sigma_{m-1}\}$ be an alphabet,
	and $m\in\{1,2,\ldots,256\}$.
	Let us explain the idea of our algorithm, denoted by $\textsf{EAH}\mbox{}n$.
	For example, let $w=w_{1}w_{2}\ldots{w_{h}}$ be a string
	over $\Sigma$. We encode $w$ by a $5$-tuple $U=(A,B,C,D,E)$, where $A,B,C,D,E\in\{0,1\}^{+}$
	are constructed as follows.
\begin{enumerate}[$\bullet$]
\item
	$\textsf{EAH}n(w).1=A.$\hspace{18pt}
	Let ${Idx}:\Sigma\rightarrow\{0,1,\ldots,m-1\}$ be a function which gets as input
	a symbol $\sigma\in\Sigma$ and outputs an index $i$ such that $\sigma=\sigma_{i}$.
	If $h\geq{n}$, then $A={b10b2}({Idx}(w_{1}))\ldots
	{b10b2}({Idx}(w_{n}))$, that is,
	$A$ is the conversion of the sequence ${Idx}(w_{1}),\ldots,{Idx}(w_{n})$ from base 10
	to base 2, and $|A|=n*\lceil\log_{2}m\rceil$. Otherwise, if $h<n$, then we consider
	$A={b10b2}({Idx}(w_{1}))\ldots{b10b2}({Idx}(w_{h}))$, that is,
	$A$ is the conversion of ${Idx}(w_{1}),\ldots,{Idx}(w_{h})$ from base 10
	to base 2, and $|A|=h*\lceil\log_{2}m\rceil$.
\item
	$\textsf{EAH}n(w).2=B=B_{0}B_{1}\ldots{B_{m^{n}-1}}$, where $B_{j}\in\{0,1\}$ is def\mbox{}ined by:
\begin{displaymath}
	B_{j}=
	\left\{
		\begin{array}{ll}
		1		& \textrm{if $\exists$ $i\in\{0,\ldots,m-1\}$ and $\exists$ $k$ such that
						$w_{k}\ldots{w_{k+n}}$}\\
				& \textrm{$={Idx}^{-1}({b10}(j,m).1)\ldots{{Idx}^{-1}({b10}(j,m).n)\cdot\sigma_{i}}$.}\\
		0		& \textrm{otherwise.}
		\end{array}
	\right.
\end{displaymath}
	for all $j$, $0\leq{j}\leq{m^{n}-1}$, and ${b10}(j,m)$ is a tuple of length $n$ denoting
	the conversion of $j$ from base 10 to base $m$, such that ${Len}({b10}(j,m))=n$ and
	${b10}(j,m).i$ is
	the $i$-th digit (from left to right) of this conversion.
\item
	$\textsf{EAH}n(w).3=C=C_{0}C_{1}\ldots{C_{m-1}}$, where $C_{i}=C_{i}^{0}C_{i}^{1}\ldots{C_{i}^{m^{n}-1}}$,
	for all $i$, $0\leq{i}\leq{m-1}$, and $C_{i}^{j}\in\{0,1,\lambda\}$ is def\mbox{}ined by:
\begin{displaymath}
	C_{i}^{j}=
	\left\{
		\begin{array}{ll}
		1	& \textrm{if $B_{j}=1$ and $\exists$ $k\in\{1,\ldots,h-n\}$ such that $w_{k}\ldots{w_{k+n}}$}\\
& \textrm{$={Idx}^{-1}({b10}(j,m).1))\ldots{{Idx}^{-1}({b10}(j,m).n)\cdot\sigma_{i}}$.}	\\
		0	& \textrm{if $B_{j}=1$ and $\nexists$ $k\in\{1,\ldots,h-n\}$ such that $w_{k}\ldots{w_{k+n}}$}\\
& \textrm{$={Idx}^{-1}({b10}(j,m).1))\ldots{{Idx}^{-1}({b10}(j,m).n)\cdot\sigma_{i}}$.}	\\
		\lambda	& \textrm{if $B_{j}=0$.}
		\end{array}
	\right.
\end{displaymath}
	for all $j$, $0\leq{j}\leq{m^{n}-1}$.
	It is clear that $|C|=m*|\{k \mid B_{k}=1\}|$.
\item
	$\textsf{EAH}n(w).4=D=D_{0}D_{1}\ldots{D_{m-1}}$, where $D_{i}=D_{i}^{0}D_{i}^{1}\ldots{D_{i}^{m^{n}-1}}$
	for all $i$, $0\leq{i}\leq{m-1}$, and $D_{i}^{j}\in\{0,1\}^{*}$ is def\mbox{}ined by:
\begin{displaymath}
	D_{i}^{j}=
	\left\{
		\begin{array}{ll}
	{Mb10b2}({Freq}(\sigma_{i},j))			& \textrm{if $C_{i}^{j}=1$.}		\\
		\lambda						& \textrm{if $C_{i}^{j}\neq{1}$.}
		\end{array}
	\right.
\end{displaymath}
	where ${Freq}(\sigma_{i},j)=|\{k\mid w_{k}\ldots{w_{k+n}}=\sigma_{b10(j,m).1}
	\ldots{\sigma_{b10(j,m).n}}\sigma_{i}\}|$.
	Let us denote by ${Marked}$ the set $\{(i,j)\mid C_{i}^{j}=1\}$.
	The greatest element of the set
	$\{|{b10b2}({Freq}(\sigma_{i},j))| \mid (i,j)\in{{Marked}}\}$ is denoted by $Max$.
	Then, ${Mb10b2}({Freq}(\sigma_{i},j))$ is def\mbox{}ined by:
\begin{displaymath}
	{Mb10b2}({Freq}(\sigma_{i},j)) =
	\underbrace{00\ldots{0}}_{t(i,j)}{b10b2}({Freq}(\sigma_{i},j))
\end{displaymath}
	where $t(i,j)=Max-|{b10b2}({Freq}(\sigma_{i},j))|$.
\item
	$\textsf{EAH}n(w).5=E$
	denotes the compression of $w$ using $A,B,C,D$, adaptive codes of order $n$, and
	the Huf\mbox{}fman algorithm.
\end{enumerate}
\begin{figure}
\setlength{\unitlength}{1pt}
\begin{picture}(335,538)
	\put(10,528){\footnotesize{$\textsf{EAH}n$(string $w=w_{1}w_{2}\ldots{w_{h}}\in\Sigma^{+}$,
		where $w_{i}\in\Sigma$ for all $i$, $1\leq{i}\leq{h}$)}}
	\put(10,518){\footnotesize{$1.$ $G_{n}(w):=(\emptyset,\emptyset,f);$
		$A:=\lambda;B:=\lambda;C:=\lambda;D:=\lambda;E:=\lambda;$}}
	\put(10,508){\footnotesize{$2.$ if $h>n$ then}}
	\put(20,498){\footnotesize{begin}}
	\put(10,488){\footnotesize{$2.1.$ for $j=1$ to $h-n$ do
		$G_{n}(w).1:=G_{n}(w).1\cup{\{w_{j}w_{j+1}\ldots{w_{j+n-1}}\}};$}}
	\put(10,478){\footnotesize{$2.2.$ for $j=n+1$ to $h$ do
		$G_{n}(w).1:=G_{n}(w).1\cup{\{w_{j}\}};$}}
	\put(10,468){\footnotesize{$2.3.$ for $j=1$ to $h-n$ do
		$G_{n}(w).2:=G_{n}(w).2\cup{\{(w_{j}\ldots{w_{j+n-1}},w_{j+n})\}};$}}
	\put(10,458){\footnotesize{$2.4.$ if $n=1$ then}}
	\put(27.5,448){\footnotesize{begin}}
	\put(10,438){\footnotesize{$2.4.1.$ for $j=1$ to $h-1$ do}}
	\put(10,428){\footnotesize{$2.4.2.$ if $w_{j}=w_{j+1}$ then
		$G_{n}(w).2:=G_{n}(w).2\cup{\{(w_{j},w_{j}\_aux),(w_{j}\_aux,w_{j})\}};$}}
	\put(27.5,418){\footnotesize{end}}
	\put(10,408){\footnotesize{$2.5.$ else for $j=n+1$ to $h-1$ do
		$G_{n}(w).2:=G_{n}(w).2\cup{\{(w_{j},w_{j-n+1}\ldots{w_{j}})\}};$}}
	\put(20,398){\footnotesize{end}}
	\put(10,388){\footnotesize{$3.$ if $n=1$ then}}
	\put(10,378){\footnotesize{$4.$ for $j=1$ to $h-1$ do}}
	\put(10,368){\footnotesize{$5.$ if $w_{j}=w_{j+1}$ then
		$G_{n}(w).1:=G_{n}(w).1\cup{\{w_{j}\_aux\}};$}}
	\put(10,358){\footnotesize{$6.$ for each $e\in{G_{n}(w).2}$ do $f(e):=(0,\lambda);$}}
	\put(10,348){\footnotesize{$7.$ for $i=n+1$ to $h$ do}}
	\put(20,338){\footnotesize{begin}}
	\put(10,328){\footnotesize{$7.1.$ if $n=1$ and $w_{i-1}=w_{i}$ then
		let $e\in{G_{n}(w).2}$ be $(w_{i},w_{i}\_aux);$}}
	\put(10,318){\footnotesize{$7.2.$ else
		let $e\in{G_{n}(w).2}$ be $(w_{i-n}\ldots{w_{i-1}},w_{i});$}}
	\put(10,308){\footnotesize{$7.3.$ $f(e).1:=f(e).1+1;$}}
	\put(20,298){\footnotesize{end}}
	\put(10,288){\footnotesize{$8.$ for each $w_{j}w_{j+1}\ldots{w_{j+n-1}}\in{Vbase}$ do}}
	\put(20,278){\footnotesize{begin}}
	\put(10,268){\footnotesize{$8.1.$ $fu:=();count:=0;$}}
	\put(10,258){\footnotesize{$8.2.$
		$tuple:=Sorted(output_{b}^{G_{n}(w)}(w_{j}
		\ldots{w_{j+n-1}})\cup{output_{a}^{G_{n}(w)}(w_{j}\ldots{w_{j+n-1}})});$}}
	\put(10,248){\footnotesize{$8.3.$ for $i=1$ to $Len(tuple)$ do
		$fu:=fu\vartriangleleft{f(tuple.i).1};$}}
	\put(10,238){\footnotesize{$8.4.$ for $i=1$ to $Len(tuple)$ do}}
	\put(27.5,228){\footnotesize{begin}}
	\put(10,218){\footnotesize{$8.4.1.$ $count:=count+1;$}}
	\put(10,208){\footnotesize{$8.4.2.$ $f(tuple.i).2:=\textsf{Huffman}(fu).count.1;$}}
	\put(27.5,198){\footnotesize{end}}
	\put(20,188){\footnotesize{end}}
	\put(10,178){\footnotesize{$9.$ $A:=b10b2(Idx(w_{1}))\ldots{b10b2(Idx(w_{min(h,n)}))};$}}
	\put(5.5,168){\footnotesize{$10.$ for $j=0$ to $m^{n}-1$ do}}
	\put(5.5,158){\footnotesize{$11.$ if
		$(\sigma_{b10(j,m).1}\ldots\sigma_{b10(j,m).n},\sigma_{i})\in{G_{n}(w).2}$ then
		$B:=B\cdot{1};$ else $B:=B\cdot{0};$}}
	\put(5.5,148){\footnotesize{$12.$ for $i=0$ to $m-1$ do}}
	\put(5.5,138){\footnotesize{$13.$ for $j=0$ to $m^{n}-1$ do}}
	\put(5.5,128){\footnotesize{$14.$ if $B_{j}=1$ then}}
	\put(20,118){\footnotesize{begin}}
	\put(5.5,108){\footnotesize{$14.1.$ if $n=1$ and $i=j$ then
		$e:=(\sigma_{i},\sigma_{i}\_aux);$}}
	\put(5.5,98){\footnotesize{$14.2.$ else $e:=(\sigma_{b10(j,m).1}\ldots\sigma_{b10(j,m).n},\sigma_{i});$}}
	\put(5.5,88){\footnotesize{$14.3.$ if $e\in{G_{n}(w).2}$ then}}
	\put(5.5,78){\footnotesize{$14.4.$ begin $C:=C\cdot{1};$ $D:=D\cdot{Mb10b2(f(e).1)};$ end
		else $C:=C\cdot{0};$}}
	\put(20,68){\footnotesize{end}}
	\put(5.5,58){\footnotesize{$15.$ for $i=n+1$ to $h$ do}}
	\put(20,48){\footnotesize{begin}}
	\put(5.5,38){\footnotesize{$15.1.$ if $n=1$ and $w_{i-1}=w_{i}$ then
		$e$:=$(w_{i},w_{i}\_aux);$ else $e$:=$(w_{i-n}\ldots{w_{i-1}},w_{i});$}}
	\put(5.5,28){\footnotesize{$15.2.$ $E:=E\cdot{f(e).2};$}}
	\put(20,18){\footnotesize{end}}
	\put(5.5,8){\footnotesize{$16.$ return $(A,B,C,D,E);$}}

	\put(2,540){\line(1,0){332}}
	\put(2,2){\line(1,0){332}}
	\put(2,540){\line(0,-1){538}}
	\put(334,540){\line(0,-1){538}}
\end{picture}
\caption{The EAH data compression algorithm.}
\end{figure}
\hspace{0pt}\indent
	If $S$ is a set of edges, then $Sorted(S)$ denotes a tuple $U$ which contains the elements of $S$, but
	sorted with a certain algorithm (the algorithm is not important here, but must be f\mbox{}ixed).
\newline\indent
	Let $w$ be an input data string.
	We denote by $\textsf{H}(w)$ the encoding of $u$ using the Huf\mbox{}fman algorithm,
	and by
	$\textsf{LZ}(w)$ the encoding of $w$ using the Lempel-Ziv data compression algorithm \cite{zl:1}.
	Let us consider the following notations.
\begin{itemize}[$\bullet$]
\item $LH(w)=|\textsf{H}(w)|$.
\item	$LLZ(w)=|\textsf{LZ}(w)|$.
\item $LEAHn(w)=\sum_{i=1}^{5}|\textsf{EAH}n(w).i|$.
\end{itemize}
\begin{example}
	Let $\Sigma=\{a,b,c,d,e\}$ be an alphabet, and $w$ a string of length 200 over $\Sigma$, given
	as below (between brackets).
\newline
	$w$=
	[\texttt{abedcababedccabedcedcababedcedcccabedcabedcedccababedc}
\newline
	\texttt{abedccccedccedccedcababedcabedcedccedcababedcabedccabedcab}
\newline
	\texttt{abedcedcccccedcabedcabedccccedcccabedcccedccabedccccabedcc}
\newline
	\texttt{ababedcabedcedccabedcababedced}].
\newline\indent
	Applying $\textsf{\textup{EAH}}1$ to the string $w$, we get the following adaptive graph of order one.
\begin{figure}[h]
\setlength{\unitlength}{1pt}
\begin{picture}(346,85)
	\put(10,75){\oval(15,15)}
	\put(10,20){\oval(15,15)}
	\put(160,75){\oval(15,15)}
	\put(160,20){\oval(15,15)}
	\put(320,75){\oval(15,15)}
	\put(320,20){\oval(30,15)}

	\put(7,72){$b$}
	\put(157,72){$a$}
	\put(317,72){$c$}
	\put(7,17){$e$}
	\put(157,17){$d$}
	\put(307,17){$c\_aux$}

	\put(17.5,72){\vector(1,0){135}}
	\put(152.5,78){\vector(-1,0){135}}
	\put(10,67.5){\vector(0,-1){40}}
	\put(17.5,20){\vector(1,0){135}}
	\put(312.5,75){\vector(-1,0){145}}
	\put(167.1,20){\line(1,0){124.9}}
	\put(292,20){\line(0,1){49}}
	\put(292,69){\vector(1,0){22}}
	\put(312.8,72){\line(-1,0){119.8}}
	\put(193,72){\line(0,-1){30}}
	\put(193,42){\line(-1,0){179}}
	\put(14,42){\vector(0,-1){15}}
	\put(323,27.5){\vector(0,1){40}}
	\put(317,67.5){\vector(0,-1){40}}

	\put(225,78){$\scalebox{1}[0.7]{\scriptsize{(22,10)}}$}
	\put(75,81){$\scalebox{1}[0.7]{\scriptsize{(31,0)}}$}
	\put(76.5,66){$\scalebox{1}[0.7]{\scriptsize{(8,0)}}$}
	\put(3,37){\rotatebox{90}{$\scalebox{1}[0.7]{\scriptsize{(23,1)}}$}}
	\put(72,45){$\scalebox{1}[0.7]{\scriptsize{(14,11)}}$}
	\put(74,23){$\scalebox{1}[0.7]{\scriptsize{(37,0)}}$}
	\put(226,23){$\scalebox{1}[0.7]{\scriptsize{(36,0)}}$}
	\put(310,38){\rotatebox{90}{$\scalebox{1}[0.7]{\scriptsize{(28,0)}}$}}
	\put(325,40){\rotatebox{90}{$\scalebox{1}[0.7]{\scriptsize{(0,$\lambda$)}}$}}

\end{picture}
\caption{$G_{1}(w)$.}
\end{figure}
\newline\indent
	It is easy to verify that we obtain:
\begin{displaymath}
	LH(w)=462\hspace{50pt}LEAH1(w)=310\hspace{50pt}LLZ(w)=388
\end{displaymath}
	which shows that in this case $\textsf{\textup{EAH}}1$ substantially outperforms
	the well-known Lempel-Ziv universal data compression algorithm \cite{zl:1}.
\end{example}

\section{Conclusions and Further Work}
	Adaptive codes associate variable-length codewords
	to symbols being encoded depending on the previous symbols in the input data string.
	This class of codes has been introduced in \cite{t:1} as a new class of non-standard variable-length codes.
	New algorithms for data compression
	based on adaptive codes of order one, have been presented in \cite{t:2},
	where we have behaviorally shown that for a large class of input data strings, these algorithms
	substantially outperform the Lempel-Ziv universal data compression algorithm \cite{zl:1}.
\newline\indent
	The \textsf{EAH} encoder has been proposed in \cite{t:3},
	as an improved generalization of these algorithms. The work carried out so far \cite{t:2,t:3}
	has behaviorally shown that \textsf{EAH} is a very promising data compression algorithm.
	In this paper, we presented a translation of the \textsf{EAH} algorithm into the graph theory.
	Further work in this area will focus on f\mbox{}inding new improvements for the \textsf{EAH} algorithm,
	as well as new algorithms for data compression based on adaptive codes.

\end{article}
\end{document}